\pdfoutput=1
\documentclass[aps,showpacs,prl,%
  reprint,
  superscriptaddress]{revtex4-1}
\usepackage{graphicx}
\usepackage{natbib}
\usepackage{amsmath}
\usepackage{epstopdf}
\usepackage{color}
\usepackage{upgreek}

\begin{document}

\newcommand{\tj}[6]{ \begin{pmatrix}
  #1 & #2 & #3 \\
  #4 & #5 & #6
 \end{pmatrix}}

\newcommand{\sj}[6]{ \begin{Bmatrix}
  #1 & #2 & #3 \\
  #4 & #5 & #6
 \end{Bmatrix}}
\newcommand{\degree}{^\circ}
\newcommand{\unit}[1]{\,\mathrm{#1}}
\newcommand{\Figureref}[1]{Figure~\ref{#1}}
\newcommand{\Eqnref}[1]{Equation~(\ref{#1})}
\newcommand{\sgn}{\textup{sgn}}

\newcommand{\affilOU}{Homer L. Dodge Department of Physics and Astronomy, The University of Oklahoma, 440 W. Brooks St. Norman, OK 73019, USA}
\newcommand{\affilStutt}{5. Physikalisches Institut, Universit\"{a}t Stuttgart, Pfaffenwaldring 57 D-70550 Stuttgart, Germany}

\title{Atom Based Vector Microwave Electrometry Using Rubidium Rydberg Atoms in a Vapor Cell}

\author{J. Sedlacek}
\affiliation{\affilOU}
\author{A. Schwettmann}
\affiliation{\affilOU}
\author{H. K\"{u}bler}
\affiliation{\affilOU}
\affiliation{\affilStutt}
\author{J.P. Shaffer}
\email[]{shaffer@nhn.ou.edu}
\affiliation{\affilOU}
\date{\today}

\begin{abstract}
It is clearly important to pursue atomic standards for quantities like electromagnetic fields, time, length and gravity. We have recently shown, using Rydberg states, that Rb atoms in a vapor cell can serve as a practical, compact standard for microwave electric field strength. Here, we demonstrate, for the first time, that Rb atoms excited in a vapor cell can also be used for vector microwave electrometry by using Rydberg atom electromagnetically induced transparency. We describe the measurements necessary to obtain an arbitrary microwave electric field polarization at a resolution of $0.5\degree$. The experiments are compared to theory and found to be in excellent agreement.
\end{abstract}

\maketitle


Quantum systems, such as atoms, have already been adopted as time and length standards because they offer significant advantages for making stable and uniform measurements of these quantities \cite{Hall2006,Gabrielse2008}. Atoms have also been successfully used for magnetometry, reaching impressive sensitivity and spatial resolutions \cite{Budker2007,Patton12,Romalis2005,Budker2010,Polzik2010,Mitchell2010}. Despite these successes, it is only recently that atoms have been used for practical microwave (MW) electrometry and achieved sensitivities below current standards by exploiting the properties of Rydberg atoms \cite{MWpaper}. Rydberg atoms have been used for electrometry for some time, but almost exclusively in elaborate laboratory setups \cite{Merkt1999,Adams1,Adams2,Martin2013,Martin2012,Weis1986,Hotop1998,Spreeuw2010,Haroche1983,Auzinsh2001,Dunning1993,Walther1980}.

The relative lag of atom based electrometry compared to magnetometry is not simply due to a lack of importance. The accurate measurement of MW electric field strength and polarization offers interesting possibilities for antenna calibration and MW electronics development, as well as for realizing an atomic candle for MW electric field stabilization \cite{camparo1998,swan2001precision}, to name a few important examples. Atom based MW electrometry, therefore, has the potential to lead to revolutionary advances in the development of MW electronics, advanced radar applications, and materials used in MW systems. So far, only the magnetic field has been accessible in the near-field MW regime \cite{Treutlein10,Treutlein12} and our method can be valuable for measuring MW electric fields in the near-field. Recall, there is not generally a straightforward relation between the MW magnetic and electric fields in the near-field.

\begin{figure}[t]
  \includegraphics[width=\columnwidth]{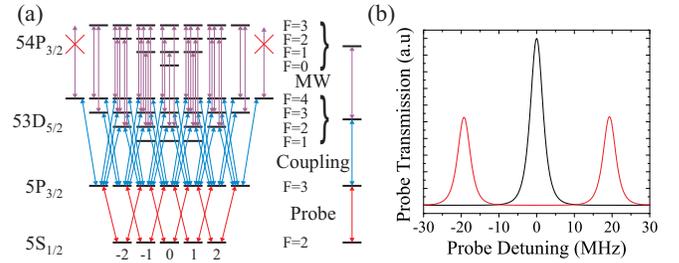}
  \caption{(Color online) (a) Level diagram showing all 52 possible states addressed by the experiment.  The arrows indicate allowed excitations for $\sigma$ polarized probe and coupling beams and $\pi$ polarized MW's.  The $54P_{3/2}$ states are shown above the $53D_{5/2}$ states for simplicity. On the right, the corresponding effective 4-Level system is shown.
  (b) Theoretical lineshapes resulting from a 3-level (black) and 4-level (\textcolor{red}{red}) system.}
  \label{fig:levels}
\end{figure}

In this paper, we demonstrate a scheme for vector MW electrometry using Rydberg atom electromagnetically induced transparency (EIT) \cite{Fleisch05,acell} in Rb atomic vapor cells. We achieve an angular resolution of $0.5\degree$ and show the method can be realized by comparing experimental data to theory. The agreement between theory and experiment is shown to be excellent. The vector measurements here are compatible with our prior work where we attained a minimum detectable electric field amplitude of $\sim 8\,\mu$V$\,$cm$^{-1}$ and a sensitivity of $\sim 30 \,\mu$V$\,$cm$^{-1}\,$Hz$^{-1/2}$ \cite{MWpaper}. To date, EIT has been principly used for vector magnetometry \cite{Cox11,Yudin10}.


To  measure the strength and polarization of a MW electric field, we use the Rb level system shown in \Figureref{fig:levels}a. In the 3-level system, $5S_{1/2}-5P_{3/2}-53D_{5/2}$, quantum interference can create a ''dark state'' that prohibits resonant absorption of a probe laser, \Figureref{fig:levels}b$\,$(black). Coupling a fourth level to this Rydberg atom EIT system, $54P_{3/2}$, with a MW electric field can create a ''bright state'' that causes probe photons to again be absorbed on resonance \cite{MWpaper,Dutta07,imamoglu1996,Lukin1999,sandhya1997}. The bright state induced by the MW electric field can manifest itself in the probe absorption spectrum as a splitting of the dark state for large enough MW electric field amplitudes, \Figureref{fig:levels}b$\,$(red). We have already shown this 4-level Rydberg atom EIT effect can be used to measure the amplitude of a MW electric field with high accuracy and sensitivity \cite{MWpaper}. In contrast to sensing only MW electric field strength, we present a significant extension of our method where, for the first time, we show it is capable of measuring the vector character, or polarization, of the MW electric field.

\begin{figure}[t]
  \includegraphics[width=\columnwidth]{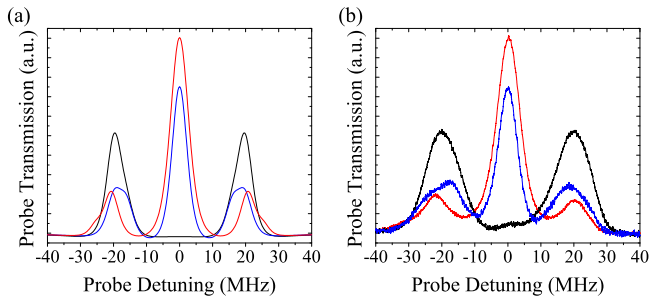}
\caption{(Color online) Theoretical (a) and experimental (b) results for the illustrative polarization cases described in the text:  probe laser, coupling laser and MWs $\hat{x}$-polarized (black); probe and coupling laser $\hat{y}$-polarized and MWs $\hat{x}$-polarized (\textcolor{blue}{blue}); probe and coupling lasers $\sigma^+$ polarized and MWs $\hat{z}$-polarized (\textcolor{red}{red}). The additional line broadening observed in the experiment is due to the MWs being inhomogeneous over the extent of the vapor cell. The effect resulted from the positioning of the antenna that was required to avoid unwanted reflections in our laboratory.}
\label{fig:theo}
\end{figure}

We exploit the hyperfine structure of the Rydberg states and the associated selection rules to measure the MW electric field polarization. EIT is known to be sensitive to the laser polarizations \cite{McGloin00,chen2000,chen2002} and we use this fact. The MW electric field polarization can be determined from the probe laser transmission by recognizing that the $53D_{5/2}$ $(F=4$ $m_F = \pm 4)$ states can be coupled or uncoupled to the $54P_{3/2}$ manifold depending on the probe and coupling laser polarization relative to that of the MW electric field. Some excitation pathways present in the system, shown in \Figureref{fig:levels}a, that pass through the stretched $53D_{5/2}$  $(F=4$ $m_F = \pm 4)$ states are restricted to the 3 levels of the EIT ladder system, $5S_{1/2}-5P_{3/2}-53D_{5/2}$. Other excitation pathways take the system through the non-stretched $53D_{5/2}$ states and can experience the full 4 level system. The behavior of the entire 52-state system, when hyperfine structure is included, can be understood by considering a few cases of laser and MW electric field polarizations that lead to 3- or 4- level behavior. \Figureref{fig:theo} shows key polarization combinations that illustrate the mixture of 3- and 4-level behavior for selected laser and MW electric field polarizations. Experimental data and theoretical results, obtained from a density matrix approach to the 52-state system, including Doppler averaging \cite{MWpaper}, are shown together for comparison. These examples illustrate the essentials of the approach.

\begin{figure}[t]
  \includegraphics[width=\columnwidth,trim=0cm 3.5cm 0cm 0cm, clip=true]{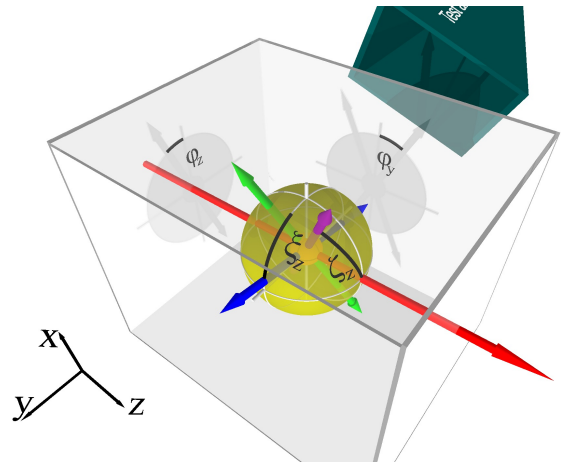}
  \caption{(Color online) Schematic view of the setup including the cell in the foreground and the test antenna in the background. The laser propagation direction (red), the polarization of the two laser beams (blue) and an arbitrary polarization direction of the MW (magenta) are shown together with the relevant angles between them as described in the main text. The shadows are the projections onto the $\hat{x}-\hat{y}$ plane on the left and the $\hat{x}-\hat{z}$ plane in the back.}
  \label{fig:polarrow}
\end{figure}

The case where the probe and coupling lasers are linearly polarized along the same direction as the MW's, $\xi=0\degree$ and $\zeta=90\degree$, where $\xi$ and $\zeta$ are defined in \Figureref{fig:polarrow} is shown in \Figureref{fig:theo}$\,$(black). In this case $\pi$-transitions are driven throughout the system and all the excitation pathways experience a 4-level system. The 3-level EIT dark state is split for the MW powers used for the experiment. The theoretical and experimental spectra have two transmission peaks separated by $\lambda_c/\lambda_p \times \Omega_{MW}$, where $\lambda_p$ is the probe $\lambda_c$ the coupling laser wavelength and $\Omega_{MW}$ is the MW Rabi frequency \cite{MWpaper}. The probe laser is absorbed on resonance.

Also displayed in \Figureref{fig:theo}$\,$(red) is the case where the probe and coupling lasers are $\sigma^{+}$ polarized and excite $\Delta m_F = +1$ transitions. The atoms are optically pumped such that the stretched states of the $5S_{1/2}$, $5P_{3/2}$ and $54D_{5/2}$ manifolds dominate the behavior. The MW electric field is polarized in the $\hat{z}$ direction. In this case, the 3-level excitation pathways are overwhelmingly favored since a $\pi$ MW transition cannot couple the stretched states to the $54P_{3/2}$ manifold. The experimental results shown support this explanation as a large probe transmission peak is observed and predicted on resonance.

If the probe and coupling lasers are both linearly polarized parallel to each other, e.g. $\hat{y}$-polarized, but orthogonal to the MW electric field polarization, e.g. $\hat{z}$-polarized, there are both 3-level excitation pathways and 4-level excitation pathways open, \Figureref{fig:theo}$\,$(blue). This more complex behavior comes from the fact that in a $\hat{z}$ atomic basis the MW electric field drives $\pi$-transitions, while the $\hat{y}$-polarized probe and coupling lasers can drive transitions throughout the $53D_{5/2}$ manifold, as they are in a superposition of $\sigma^+$ and $\sigma^-$ polarizations in the $\hat{z}$ basis. The experimental and theoretical spectra show reduced probe transmission on resonance and 2 probe transmission peaks split by $\lambda_c/\lambda_p \times \Omega_{MW}$.

Any MW electric field can be split into a component that couples atoms to the $54P_{3/2}$ state and one that does not. The relative strength of the components only depends on the MW electric field polarization relative to the polarization and propagation direction of the probe and coupling laser beams. When rotating parallel, linear probe and coupling laser polarizations around their propagation axes, the projection of the MW electric field on the probe and coupling laser polarization changes. The change of the MW electric field polarization projection relative to the probe and coupling laser polarization results in a variation of the probe laser transmission on resonance, and of the probe transmission spectra, in general. The probe laser transmission changes can be used to determine the MW electric field polarization since the probe and coupling laser polarizations are known. The splitting of the peaks indicative of the 4-level behavior remains relatively constant because this is largely determined by the electric field amplitude that the atoms experience and can be used to find the amplitude of the MW electric field in conjunction with the polarization measurement.

The geometry needed to describe a measurement of the MW electric field polarization is shown in \Figureref{fig:polarrow}. The incident MW electric field vector $\vec{E}$ forms an angle $\zeta_z$ with the space fixed propagation direction of the probe laser chosen to lie along the $\hat{z}$~axis. $E_z$ is the projection of the MW electric field on $\hat{z}$. The perpendicular component of the MW electric field, $\vec{E}_{\perp z}$, forms an angle $\xi_z$ with the polarization vector of the probe and coupling laser beams in the $\hat{x}$-$\hat{y}$ plane. The angle $\varphi_z$ between the $\hat{x}$~axis and $\vec{E}_{\perp z}$ can be determined by rotating the probe and coupling laser polarizations. In this case, the configuration is periodically changing from the case where $\vec{E}_{\perp z}$ and the laser fields are parallel, \Figureref{fig:theo}$\,$(black), to the case where the MW and laser fields are orthogonal, \Figureref{fig:theo}$\,$(blue). For simultaneous rotation of the probe and coupling laser beam polarizations about $\hat{z}$, $\xi_z$, \Figureref{fig:polarrow}; the probe transmission on resonance will oscillate between a minimum for $\xi_z= 0\degree$ , \Figureref{fig:theo}$\,$(black), and a maximum for $\xi_z=90\degree$, \Figureref{fig:theo}$\,$(blue). The amplitude of this oscillation measures $\zeta_z$, the projection angle along $\hat{z}$, since with increasing $\zeta_z$, $\vec{E}_{\perp z}$ increases. Measuring $\varphi$ and $\zeta$ along all three cartesian coordinate axes reveals the following:
\begin{subequations}
  \begin{align}
    \varphi_z=\tan^{-1}\left(\frac{E_x}{E_y}\right), \qquad
      \zeta_z&=\tan^{-1}\left(\frac{\left|E_{\perp z}\right|}{\left|E_z\right|}\right),
      \\
    \varphi_x=\tan^{-1}\left(\frac{E_y}{E_z}\right), \qquad
      \zeta_x&=\tan^{-1}\left(\frac{\left|E_{\perp x}\right|}{\left|E_x\right|}\right),
      \\
    \varphi_y=\tan^{-1}\left(\frac{E_x}{E_z}\right), \qquad
      \zeta_y&=\tan^{-1}\left(\frac{\left|E_{\perp y}\right|}{\left|E_y\right|}\right).
  \end{align}
\end{subequations}
The information obtained from measuring the three angles $\varphi_i$ is sufficient to determine the MW electric field polarization. The magnitude of the MW electric field can be obtained from the splitting of the transmission peaks observed as a consequence of the 4-level behavior. It is important to note that it is impossible to distinguish the angle $\zeta_i$ from $180\degree-\zeta_i$ (magenta and green arrow in \Figureref{fig:polarrow} for $i=z$), because these two cases differ only in the relative phase between $E_i$ and $E_{\perp i}$. However, as the information from $\zeta_i$ is available from a measurement of the $\varphi_i$ it can be used to improve the accuracy of the experimental measurement and serve as a self consistency check of the resonant probe transmission.

Measuring $\varphi_i$ along two of the three axes, e.g. $\varphi_z$ and $\varphi_y$, is theoretically sufficient to determine the MW electric field polarization, except for the case where the MW electric field polarization lies directly in the plane orthogonal to the 2 measurement axes ($E_x=0$). In this case, the phase of $E_z$ relative to $E_y$ cannot be determined by referencing both to $E_x$. $\zeta_z$ provides no additional information. As one would expect, a measurement out of the plane determined by the 2 measurement axes is required. Additionally, the measurement becomes sensitive to noise if $E_x$ is small and only  $\varphi_z$ and $\varphi_y$ are used to determine the MW electric field polarization. As a consequence, it is best to measure $\varphi_x$, $\varphi_y$ and $\varphi_z$, provided no a priori knowledge of the MW electric field polarization is known.

\begin{figure}[t]
\includegraphics[width=\columnwidth]{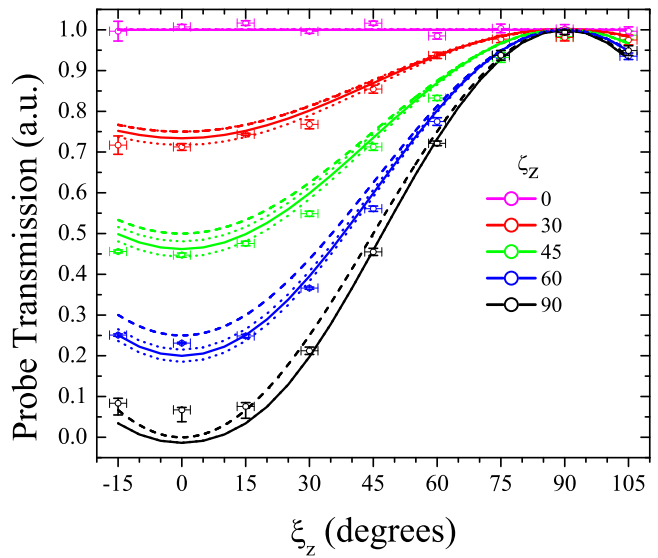}
\caption{(Color online) Probe laser transmission on resonance for different angles between the laser polarizations and the MW electric field vector. The vertical error bars for the experimental points are due to statistical errors in the measured peak height while the horizontal error bars are due to systematic uncertainty in $\xi$.  52-state theoretical results (solid lines) with $\pm 1\degree$ uncertainty in $\zeta$ (dotted lines). The curves for $1-(E_\parallel/|\vec{E}|)^2=1-\cos^2(\xi)\sin^2(\zeta)$ are also shown (dashed lines). The deviation of the \emph{}points with larger $\zeta$ around $\xi = 0 \degree$ is due to polarization impurities in the laser and MW beams.}
\label{fig:rotpol}
\end{figure}

A simplified model of the resonant probe transmission dependence on $\xi_i$ can be obtained by considering the projection of the MW electric field vector $\vec{E}$ onto the probe and coupling laser polarization. From \Figureref{fig:polarrow}, the projection of the MW electric field on the laser polarization direction is $E_\parallel=|\vec{E}|\cos(\xi)\sin(\zeta)$. Due to branching between the 3- and 4-level behavior, the resonant probe transmission can be approximated as $T=1-(E_\parallel/|\vec{E}|)^2=1-\cos^2(\xi)\sin^2(\zeta)$. The approximation neglects optical pumping effects that occur in the full 52-state system and therefore does not reproduce the probe transmission amplitude very well. However, the angular positions of the minima and maxima as $\xi_i$ is varied are predicted accurately. The simplified model yields the correct MW electric field polarization from a measurement of the $\varphi_i$. To get the correct probe transmission amplitude, the full 52-level theory including optical pumping and Doppler averaging as described in \cite{MWpaper} has to be used, \Figureref{fig:rotpol}.

The experimental setup, \Figureref{fig:polarrow}, consists of a probe laser beam and a coupling laser beam that are overlapped and counter-propagate through a cuboidal atomic Rb cell with dimensions (10$\,$mm$\times$10$\,$mm$\times$30$\,$mm). The probe laser is an extended cavity diode laser (ECDL) at $\sim 780 \unit{nm}$ that propagates along the $\hat{z}$-axis. The coupling laser is derived from a home-built frequency doubling system operating at $\sim 480 \unit{nm}$ and propagates along $-\hat{z}$. The doubled light is generated from an amplified ECDL at $960\,$nm. The probe laser is locked to the $^{87}$Rb $5S_{1/2}(F=2) \rightarrow 5P_{3/2}(F=1,3)$ crossover peak. The $960\,$nm ECDL is locked to a Fabry-Perot cavity that is stabilized to an EIT signal generated in a separate vapor cell. The laser linewidths are $\sim 700\,$kHz. An acousto-optic modulator (AOM) is used to scan the probe laser frequency around the  $5S_{1/2}(F=2) \rightarrow 5P_{3/2}(F=3)$ transition. An intensity stabilization circuit based on an FPGA \cite{schwett11} is used to intensity stabilize the probe laser to $\sim 0.1\%$. The polarizations of the laser beams are adjusted and filtered using waveplates and Glan laser polarizers. The probe (coupling) laser spot size is $200\,(65)\,\mu$m and the power is $15\,\mu$W$\,(11\,$mW). The corresponding probe (coupling) Rabi frequency is $2 \pi \times 8.1 (2 \pi \times 3.4)\,$MHz.

MW's are generated at $14.233\,$GHz with a signal generator (HP8340B). The MWs are coupled into a horn antenna that illuminates the Rb vapor cell. The propagation direction of the MW electric field relative to the lasers is shown in \Figureref{fig:polarrow}. The polarization of the MW's is linear and it is changed in the experiment by rotating the antenna. The MW intensity is calculated from the geometry and coupling efficiencies of the MW components and confirmed by comparing the associated MW electric field strength to the splitting of the probe laser transmission peaks observed in the experiment. The transition dipole moment for the transition between the Rydberg states is calculated to be $4103\,$Debye \cite{Beterov2011}. The MW intensity used for the experiments was $1.27 \times 10^{-3}\,$mW$\,$cm$^{-2}$. This corresponds to a Rabi frequency of $2 \pi \times 64\,$MHz. For these parameters the 52-state theory yields a probe transmission peak splitting of $39.36\,$MHz. The probe transmission peak splitting observed in the experiment is $39.36\pm 0.06\,$MHz, \Figureref{fig:theo}b$\,$(black).

The intensity of the coupling laser is modulated at $40\,$kHz with an AOM and the probe transmission is detected on a photodiode.  The photodiode signal is processed using a lock-in amplifier. The experimental data is a result of 20 averages. Three pairs of orthogonal Helmholtz coils surround the cell to cancel the background geomagnetic field to a level of $< 0.1\,$G. The experiment is conducted at a Rb vapor cell temperature of $45\degree$C. The temperature corresponds to a Rb vapor pressure of $2.6 \times 10^{-6}\,$Torr which remains fixed throughout the experiments. The cell was heated to prevent significant condensation of Rb on the walls of the vapor cell. Condensation of Rb on the walls of the vapor cell causes reflections of the MWs leading to spurious signals.

\Figureref{fig:rotpol} shows an example of a measurement used to determine the MW electric field polarization. The polarizations of the probe and coupling lasers are in a lin$\parallel$lin configuration. The pump and probe polarizations are rotated through $\xi_z = 120\degree$ for different MW antenna angles, $\zeta_z$. The transmission of the probe laser on resonance is plotted in the figure as a function of $\xi_z$ for different $\zeta_z$. Also shown, for comparison, is the simple model discussed previously (dashed line), and the result of the full 52-state theory (solid line). The direction of $\vec{E}_{\perp z}$ can be found because the on resonance probe transmission is minimum for probe and coupling laser polarizations parallel to $\vec{E}_{\perp z}$. The angle $\zeta_z$ between the MW electric field polarization and the laser propagation axis is determined by the modulation depth of the on resonance probe transmission as a function of $\xi_z$.

The maximum sensitivity is obtained when the 4-Level peaks are completely split from the 3-Level peak.  For our experimental parameters this occurs at a MW electric field amplitude of $\sim 10 \unit{mV\,cm^{-1}}$. Increasing the MW electric field strength has little effect on the central peak until $\sim 100\unit{mV\,cm^{-1}}$. At this point, the peak starts to shift and decrease in height, most likely due to multi-photon transitions \cite{MWpaper}.  The angular resolution detected in the experiment is $\sim 0.5\degree$ in both $\zeta$ and $\varphi$. Narrower laser linewidths, lower noise electronics, purer polarizations and better laser intensity stabilization can significantly improve the sensitivity.

In summary, we have demonstrated an atom based method for sensitively measuring the polarization of a MW electric field by making use of Rydberg states in a Rb vapor cell. The vector electrometry described here is compatible with measurements of the electric field amplitude as presented in our earlier work \cite{MWpaper} and is therefore practical for making atom based measurements of the MW electric field in compact portable setups. We were able to achieve an angular resolution of $\sim 0.5\degree$ in both $\zeta$ and $\varphi$. Our approach allows for miniaturization on the $\unit{mm}$ or even sub-mm scale \cite{MicroCellPaper}. Due to the optical readout and materials used, distortion of the electric field is relatively small compared to dipole antennas.

We thank T. Pfau and Robert L\"{o}w for useful discussions. This work was supported by the DARPA Quasar program through a grant through ARO (60181-PH-DRP) and the NSF (PHY-1104424).

\bibliography{mybib2}

\end{document}